# Quantum Phase Extraction in Isospectral Electronic Nanostructures

Christopher R. Moon,[1] Laila S. Mattos,[1] Brian K. Foster,[2] Gabriel Zeltzer,[3] Wonhee Ko,[3] Hari C. Manoharan[1*]

[1]*Department of Physics, Stanford University, Stanford, CA 94305, USA*
[2]*Department of Electrical Engineering, Stanford University, Stanford, CA 94305, USA*
[3]*Department of Applied Physics, Stanford University, Stanford, CA 94305, USA*

*To whom correspondence should be addressed.  E-mail: manoharan@stanford.edu

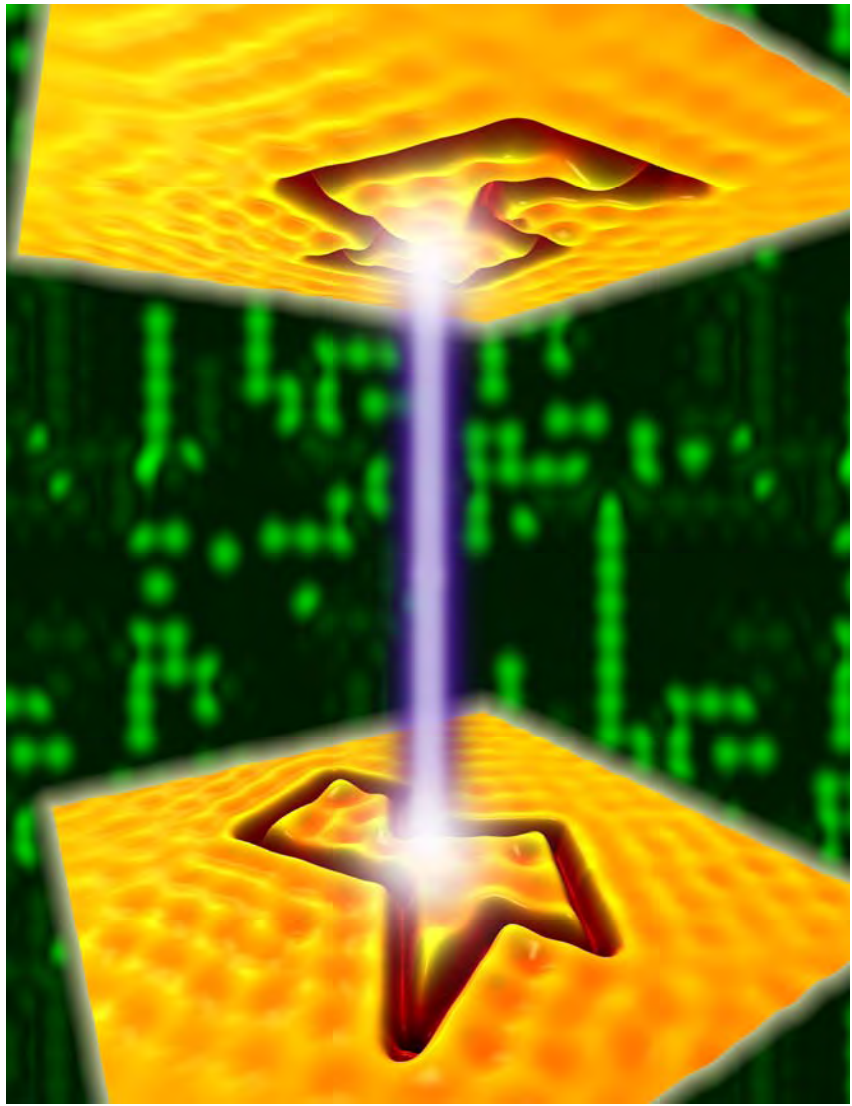

**Quantum phase is not a direct observable and is usually determined by interferometric methods. We present a method to map complete electron wave functions, including internal quantum phase information, from measured single-state probability densities. We harness the mathematical discovery of drum-like manifolds bearing different shapes but identical resonances, and construct quantum isospectral nanostructures possessing matching electronic structure but divergent physical structure. Quantum measurement (scanning tunneling microscopy) of these "quantum drums" [degenerate two-dimensional electron states on the Cu(111) surface confined by individually positioned CO molecules] reveals that isospectrality provides an extra topological degree of freedom enabling robust quantum state transplantation and phase extraction.**

The local structure of wave functions can now be experimentally determined in many materials. However, just as bonding in molecules or conductivity in solids depends not only on the magnitude of the orbital wave functions but also on the phase, the electronic properties and dynamics of nanostructures critically depend on determining both magnitude and phase of wave functions.

In a classical system, measuring the phase of a standing wave is trivial, but the internal phase of a quantum wave function $\psi$ is not an observable; only the probability density $|\psi|^2$ can be determined from direct measurement. To determine phase, some form of additional information is required.

Recently, quantum phase has been inferred in several experiments, including spectral interferometry of Rydberg wave packets (*1*) and tomography of molecular orbitals (*2*). Such experiments are performed on atoms or molecules in the gas phase and require ultrafast measurements of the response of a quantum state to an impinging wave or excitation. In solid-state systems, phase-sensitive measurements have been performed in quantum dots and rings (*3*) and with computational post-processing of diffraction patterns (*4*). All of these experiments retrieve the underlying phase differences through interferometry once a reference beam and geometry are defined.

Here we present a non-interferometric method to map the internal quantum phase of a solitary, time-independent state by harnessing the topological property of isospectrality as the additional degree of freedom. We create a particular pair of geometric shapes with matching spectral properties. We then show that the complete phase information of wave functions in both structures can be experimentally determined. Because this technique is based on a fundamental mathematical symmetry, it is general and should apply to several types of nanostructures and materials, but we demonstrate it for wave functions of confined regions on metal surfaces. To develop this method, we have drawn from recent answers to a famous, long-standing question in mathematics.

A general method for obtaining an excitation spectrum from a given geometry is to apply the Laplacian operator to the oscillation variable. Much scrutiny has been given to the relation, familiar to any musician, between the allowed frequencies of a wave—the eigenvalues of the Laplacian—and the shape of the boundary that encloses it. Given all of the eigenfrequencies of a resonator, its area can be determined (*5*), as well as its perimeter and the number of embedded holes (*6*). However, does a resonator's eigenspectrum uniquely specify its geometry? In other words, as Kac asked 40 years ago (*7*), "Can one hear the shape of a drum?"

When Kac's seminal paper was presented in 1966, Milnor had already proven the existence of two noncongruent 16-dimensional flat tori in which the Laplacian's eigenvalue spectrum is identical (*8*). In subsequent decades, mathematicians struggled to find such isospectral geometries in lower dimensional systems. Sunada discovered a group-theoretic method for proving isospectrality (*9*) in the 1980s, and Buser proved a class of isospectral manifolds by "pasting" two-dimensional (2D) flat tiles together in higher dimensions (*10*). However, the dimensionality of proven isospectral domains was only slowly whittled down to three dimensions (*10, 11*) [ironically, the one-dimensional analog of Kac's question is trivial—one can always hear the length ("shape") of a string by identifying its fundamental frequency].

Did Kac's 2D drums mimic their closest cousins in 3D, where isospectral pairs had been found, or match the 1D case where the spectrum uniquely defines the geometry? It was not until 1992 that Gordon, Webb, and Wolpert mathematically discovered the first 2D isospectral domains (*11*), finally answering Kac's enigma: No, one cannot hear the shape of a drum.

This result in ostensibly obscure mathematics crossed over to general audiences because of its connection with simple ideas of sound and shape (*12, 13*) (SOM text). However, there were still many questions about whether this theoretical spectral equality could be observed in real, imperfect systems. Sridhar and Kudrolli (*14*) performed the first experimental test of classical isospectrality by measuring resonances in microwave transmission through thin metal cavities. Even and Pieranski (*15*) verified isospectrality in actual vibrating drums made from liquid crystal films.

Because the time-independent Schrödinger equation is also a wave equation defined by the Laplacian and boundary conditions, systems of quantum particles can also theoretically be isospectral. Indeed, such systems—an electron resonator, for example—can have interesting properties impossible to study in classical drums. In fact, it has recently been argued that quantum effects can allow one to distinguish between structures that otherwise would be isospectral (*16*).

To explore quantum isospectrality and its sensitivity to our nanofabrication abilities, it is necessary to find a system whose frequency response is highly sensitive to geometry, but whose particle-particle interactions remain relatively constant. Here we detail studies that establish the limits of quantum isospectrality in a system where the 2D electron density is sufficiently high that Coulomb interactions are minimized and further screened by bulk electrons (*17*), and quasiparticle lifetimes are sufficiently long to preserve quantum coherence. We then apply these isospectral geometries to quantum phase measurement.

We designed quantum-mechanical isospectral electron resonators by following symmetry and tiling rules from the mathematical literature (*11, 18, 19*). For our "vibrating" medium, we utilized the 2D Fermi sea of electrons that inhabits the Cu(111) surface state (*20*). These electrons were confined to theoretically isospectral shapes with walls made from CO molecules positioned with the tip of a scanning tunneling microscope (STM) (*21, 22*).

We chose CO because it is easily and accurately manipulated, yet stable over a wide range of sample voltage $V$ and tunnel current $I$ (*23*). We can stably pack CO molecules closer than adsorbed atoms, and thus form nearly continuous walls (see Fig. 1, A to C). For these experiments, we operated a home-built STM at 4.2 K in ultrahigh vacuum (*13*). CO bonds directly above Cu atoms and images as a ~50 pm topographical depression at low biases. After ~0.012 monolayers of CO were dosed, the molecules were positioned to form the complex boundaries required for isospectral geometries (*13*). We then performed successively stricter tests of isospectrality: spectral matching, amplitude matching, and the transplantation of wave functions. The results are then used to map quantum phase.

The most basic known planar isospectral domains (*11*) are nine-sided polygons that have been previously termed "Bilby" and "Hawk" (*24*). Each is a polyform composed of seven identical triangles and created via a specific series of reflections, such that every triangle is the mirror image of its neighbors. As there is proven flexibility in choice of interior angles for the base triangle (*18*), and the Cu(111) surface is a close-packed triangular lattice, we chose tilings based on the 30°-60°-90° triangle. The molecular designs superimposed over the ideal Bilby and Hawk shapes and the STM topographs of the experimentally realized quantum resonators are shown in Fig. 1, A and B. Each structure was assembled from 90 individual CO molecules, bounds an area of ~57 nm$^2$, and holds ~30 electrons (*13*).

We also carried out control experiments where we violated the construction principles [SOM text (*13*)] of the isospectral structures. One structure, "Broken Hawk," was created by flipping two sections of Hawk (along yellow dashed lines, Fig. 1C). The area, perimeter, and molecule count of Broken Hawk are identical to Hawk and Bilby.

We performed *dI/dV* spectroscopy throughout each of these structures to measure their eigenenergies (*13*). In order to observe all eigenmodes, we acquired dense sets of spectra along an interior triangular lattice (white crosses in Fig. 1, A to C; 46 spectra for each structure). To graphically compare the overall spectral content of the three structures, we condense the total 138 spectra into "spectral fingerprints" as shown in Fig. 1, D to F. Here, for each structure, the *dI/dV*-vs-*V* point spectra are sorted by energy of largest peak and plotted with *dI/dV* mapped to the color scale shown. Although there are gross features in common among the structures (as expected, since all have same area and perimeter), the detailed features reveal that Bilby and Hawk are most closely matched, while Broken Hawk visibly differs. To further collapse the data set, we plot the average *dI/dV* (Fig. 1G) for each structure. Again, we find excellent agreement between the Bilby and Hawk shapes, and a significant deviation in Broken Hawk. For example, the first peak (which arises from the first two eigenmodes of each structure, as we will show below) has been shifted by ~7 mV in the Broken Hawk and the subsequent structure shows further departures. These data provide initial evidence of quantum isospectrality in suitably assembled geometries.

The natural resonance frequencies of 2D electrons in these nanostructures are on the order of $v_F / l \sim 100$ THz, where the Fermi velocity $v_F \sim 10^8$ cm/s and characteristic structure size is $l \sim 10$ nm. For reference, we have added this frequency scale to Fig. 1, D to G. As another simple yet compelling demonstration of quantum isospectrality, we have converted the measured average spectra in Fig. 1G to audio frequencies so that one can "listen" to each of the structures. The results [see Movie S1 and Methods (*13*)] portray the THz ringing of electrons as one might hear them. Indeed, Bilby and Hawk—as quantum drums—"sound the same," while Broken Hawk can be audibly distinguished.

Thus, we have been able to observe evidence for isospectrality in a quantum system even with nonideal boundaries. In an effort to quantify the limits of isospectrality in these quantum structures, we devised a complementary analysis of the ensemble data set.

In an infinite hard-wall potential, and if the confined electrons had very long lifetimes, our spectra would show sharp spikes at the eigenenergies. However, several factors widen these peaks into broader resonances. Electrons can tunnel across the confinement potential imposed by the molecular walls (*25*) or scatter from the molecules into bulk states (*26*). There is also intrinsic lifetime broadening of the surface states caused by electron-electron and electron-phonon interactions that dominate for electrons at low energies (*17*).

To overcome these linewidth effects, we applied a self-consistent fitting procedure that simultaneously matches a series of Lorentzians to all of the spectra taken in a structure [see (*13*)]. The fitted eigenenergies for Bilby and Hawk (Fig. 2A) are remarkably similar, confirming their isospectrality in our quantum system. By plotting the amplitudes of each Lorentzian across the isospectral domains (Fig. 2C, Bilby, and Fig. 2D, Hawk), the spatial structure of the underlying eigenmodes is revealed, presenting a purely experimental confirmation that the fitted energies correspond to actual quantum resonances.

All possible two-way correlations between the three data sets of Bilby/Hawk/Broken Hawk (Fig. 2A) show that Bilby and Hawk are numerically closest by a factor of ~9 in root-mean-square (rms) energy difference. To quantify variation from perfect isospectrality we plot (Fig. 2B) the deviation $\Delta V$ from the average Bilby/Hawk energy at each mode. Here it is evident that statistically Bilby and Hawk are isospectral within measurement error and Broken Hawk is distinct. As a conservative bound on quantum isospectrality in this system, we can say that that Bilby and Hawk are isospectral to better than $\pm 2\,\text{mV}$ (Fig. 2B, dashed lines), and the alterations built into Broken Hawk take it well outside this window. These tests demonstrate the extreme sensitivity exhibited by quantum isospectrality to correct geometry and gentle topological perturbations. Results for other geometric perturbations are shown in Fig. S4 (*13*).

Next, we studied isospectral quantum resonators with greater complexity (*18*), consisting of 21 triangles each (Fig. 3, A and B). We refer to these structures as "Aye-aye" and "Beluga" (or *A* and *B* in shorthand). In addition to being isospectral, this pair of structures theoretically possesses points that are homophonic; that is, not only will the same frequencies result if the drums are "struck" at these points, but the relative amplitudes of those frequencies will be identical (*27*). The homophonic points occur at the meeting of six triangles in the interiors of the two shapes.

Our experimental constructs were assembled from 105 CO molecules and hold ~14 electrons. We extracted the eigenenergies of these structures after dense spectroscopy (29 locations each) and performed the same Lorentzian-fitting procedure described above, and found them to be isospectral within experimental error ($\pm 3$ mV). Here we focus on the special aspect of these geometries: the response at the homophonic points (white crosses in Fig. 3, A and B). The spectra (Fig. 3C) show that the electronic structure at these locations is nearly identical, despite their entirely different environments.

This agreement should be contrasted with the difference observed in any other pair of isolated *dI/dV* traces acquired from isospectral resonators. For example, we measured the rms difference between each pair of *dI/dV* traces in *A* and *B*. Of the 406 distinct combinations, the minimum difference (the maximum correlation) occurs between the homophonic points in *A* and *B*; the next closest pair has twice this difference. Remarkably, if the geometry of *A* is subtly broken, the homophonic point spectrum changes far more than when the geometry is radically changed to the *B* shape [see Fig. S5 (*13*)]. Our results are highlighted in Movie S2 (*13*), which plays the audio conversion [see Methods (*13*)] of three representative pairs of point spectra, including the homophonic pair. Here, the similarity between the homophonic points is audibly contrasted with the easily discernable differences between other points. In analogy to two differently shaped drums struck exactly at homophonic points, this quantum version demonstrates how to materialize identical local elec-tronic structure at two remote locations surrounded by different global environments.

Having verified the equivalence of the electron energies in specific pairs of quantum nanostructures, we now turn to their eigenfunctions. The mathematical underpinnings of the proof of isospectrality rely on Berard's generalization (*28, 29*) of Sunada's theorem (*9*), which established a relationship between the normal modes of these shapes. Specifically, if an eigenfunction of one member of an isospectral pair is known, there exists a non-isometric transformation (*30*) yielding the corresponding eigenfunction of its isospectral complement. The transformation works by cutting the wave function data for one domain into its constituent triangles, which are then combined by appropriate superpositions and transplanted onto the second domain [SOM text (*13*)]. For example, in the 21-triangle homophonic structures the triangular sections of the wave functions are labeled (Fig. 3, A and B) and represented as column vectors, so that $\vec{A} = (A_1, A_2, \ldots, A_{21})$ and $\vec{B} = (B_1, B_2, \ldots, B_{21})$. Then, the transplantation operation is an isomorphism that can be represented by a $21 \times 21$ matrix $\mathbf{T}$ (*13, 18*), and $\vec{B} = \mathbf{T}\vec{A}$. Every row of $\mathbf{T}$ has 5 non-zero elements, meaning that every triangle of a *B* wave function is the superposition of 5 sections of the *A* wave function. Every non-zero element of $\mathbf{T}$ is $\pm 1$, where the negative sign is accompanied by flipping the triangle about one of its sides to match the symmetry of its destination triangle. We note that $\mathbf{T}$ is energy independent, and hence applies to any wave function. Incidentally, there also exists a second (linearly independent) transplantation matrix, $\mathbf{T}'$, which has 16 non-zero elements per row; therefore any norm-preserving linear combination of $\mathbf{T}$ and $\mathbf{T}'$ will also yield valid transplantations.

To measure the wave functions of the homophonic structures, we acquired high-resolution open-loop *dI/dV* maps at each mode energy [see (*13*)]. We start with the experimental measurements of the ground state mode in *A* and *B* (Fig. 4A). When the *A* mode is then transplanted, the result is in excellent agreement with the data for the *B* mode. Concordantly, by inverting the transplantation matrix $\mathbf{T}$, the *B* data can be transformed into a match for the *A* data. We have also tested the $\mathbf{T}'$ matrix described above with similar results, and we emphasize that the structure of the matrices afford no fitting parameters as they are exactly determined by geometry and symmetry. As the original mathematical proof of isospectrality employed transplantation, this experimental observation is arguably the strongest confirmation of quantum isospectrality in our nanostructures. Perhaps the most tantalizing component of transplantation, however, appears when addressing the excited states. As is, their transplantation fails utterly [Fig. S6 (*13*)] because we require a final crucial ingredient: the internal, quantum-mechanical phase of the wave functions.

We recognized that our quantum invocation of Sunada's theorem is inherently phase-sensitive because it must act on $\psi$ rather than $|\psi|^2$. Hence, our basic idea was to determine the phase of measured wave function data inside one isospectral domain by optimizing its transplantation to the data for its partner domain. That is, we wished to minimize the error $\delta = \sum_{\mathbf{r}} \left| \left\| \mathbf{T} | A(\mathbf{r}) \rangle \right\|^2 - \left\| B(\mathbf{r}) \rangle \right\|^2 \right|$, summing over every data point in real space $\mathbf{r}$. Here, $\left\| A \rangle \right\|^2$ and $\left\| B \rangle \right\|^2$ are the measured probability densities determined from our *dI/dV* maps (*13*). In the homophonic structures (Fig. 3, A and B), we measured these probability densities for the first few eigenmodes at ~$10^5$ points each. Separating *A* into its amplitude (known) and phase (unknown), $|A\rangle = \left\| A(\mathbf{r}) \rangle \right\| e^{i\phi(\mathbf{r})}$, we designed and operated a "quantum transplantation machine" (QTM) to extract the optimal

phase $\phi(\mathbf{r})$ by minimizing $\delta$ (*13*). The QTM [Fig. S3 (*13*)] has two inputs, two outputs, and consists of simple operations surrounding the transplantation operation **T** [Eq. S1, (*13*)].

In this parameterization of the wave function, the magnitude is continuously varying, so the internal phase (modulo an overall arbitrary phase factor) can be expressed as a set $\phi(\mathbf{r}) \in \{0, \pi\}$ for all $\mathbf{r}$. That is, the quantum phase difference between any two points in a single wave function is a well-defined quantity and can only be 0 or $\pi$. This phase, or equivalently the sign information of the wave function $e^{i\phi(\mathbf{r})} = \pm 1$, is normally irretrievably obscured by quantum measurement of a single wave function. The importance of $\phi(\mathbf{r})$ stems from its crucial role in superpositions and dynamics, as this phase evolves in time according to the time-dependent Schrödinger equation.

Determination of these internal signs of a wave function is a type of inverse problem in the class of binary optimization. To solve it, we began by assigning a random phase at every point; at this stage, $|A\rangle$ and $\mathbf{T}|A\rangle$ were speckled and irregular (see Fig. 4B, initial QTM state for mode 3). Next, the phase was adjusted iteratively to minimize $\delta$ using a greedy algorithm (*13*, *31*). Figure 4C shows an intermediate state of the QTM while phase extraction is still in progress. Upon convergence, the final output of the QTM (Fig. 4D) is not only $|A\rangle$, but also $\mathbf{T}|A\rangle = |B\rangle$; that is, the full phase information of the wave functions of both isospectral structures has been determined. Similarly, the mode 2 and 1 electron wave functions extracted for both the *A* and *B* resonators are displayed in Fig. 4, E and F, respectively.

Movie S3 (*13*) presents the full evolution of the QTM for each of these modes as the phase is extracted. The convergence of the QTM is quite robust, thanks to the acute phase sensitivity of transplantation. Time-reversal symmetry forces our wave functions to be real; hence, the only additional degree of freedom allowed in the phase extraction is an overall $\pi$ phase shift, or sign change. In fact, we observe this vividly in the statistics of the extraction process: the QTM, for random initial phase seed, converges with 50% probability to each of two possible outcomes linked by a global $e^{i\pi}$ phase factor. An example can be seen by comparing the QTM results for mode 2 in Movie S3 with Fig. 4E.

Our work shows that phase extraction is surprisingly robust even in the presence of noise and imperfect boundary conditions—indicating accessibility in other systems—and we note that other realizations can be even simpler [see Fig. S1 (*13*)] when not constrained by a lattice as in our experiment. The application of fundamental geometry to quantum systems is a growing field (*32*, *33*). Quantum phase extraction using topological symmetries has general applicability to the fields of quantum dots, nanoscale devices, and molecular electronics. The necessary criterion is geometric control on the scale of the relevant wavelength, already attainable in technologies like nanolithography, self-assembly, and molecular design. Quantum dots, suitably patterned in semiconductors or metals, should mimic the physics described here, and calculations exist as a guideline for building isospectral systems (*34*). Measurements of nanomechanical resonators, now approaching the quantum limit (*35*), will similarly be phase-obscured and can benefit from these methods.

In chemistry, a complementary quest exists for isospectral and near-isospectral molecules, which can theoretically exist for a variety of different potentials but have not been experimentally observed (*36*, *37*). Just as this work allows the STM to be used as a scanning phase meter beyond charge sensitivity, we envision that other phase-sensitive experimental probes can be obtained by similar geometric tuning of quantum materials.

38. Supported by the NSF (CAREER DMR-0135122 & IMR DMR-0216913), DOE (DE-AC02-76SF00515), ONR (YIP/PECASE N00014-02-1-0351), the Research Corporation (RI0883), and the Stanford-IBM Center for Probing the Nanoscale (NSF PHY-0425897). We acknowledge the NDSEG program (C.R.M. & B.K.F.) and the Alfred P. Sloan Foundation (H.C.M.) for fellowship support. We thank R. E. Schwartz, S.-H. Song, A. C. Manoharan, J. T. Moon, D. P. Arovas, M. Zworski, and M. R. Beasley for discussions, and R. G. Harris for expert technical assistance.**Supporting Online Material**

www.sciencemag.org

SOM Text

Materials and Methods

Figs. S1 to S6

Movies S1 to S3

References

– 9 –

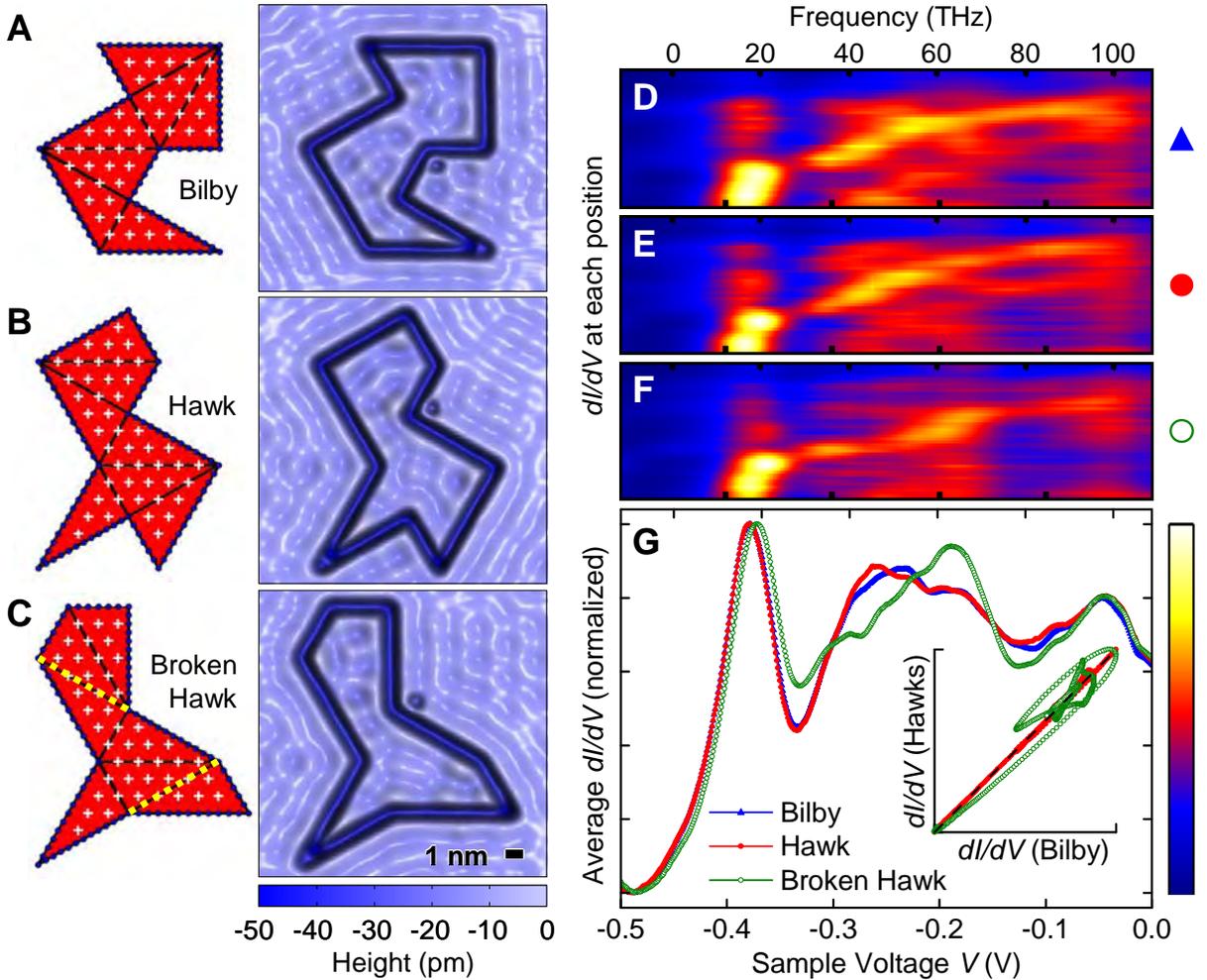

**Fig. 1 | Design and realization of quantum isospectral resonators assembled from CO molecules on the Cu(111) surface.** Schematics and STM topographs of the (**A**) Bilby, (**B**) Hawk, and (**C**) Broken Hawk domains. The seven identical 30-60-90 triangles composing each shape are shown in red. Blue dots indicate the positions of wall molecules. White crosses mark locations where $dI/dV$ spectroscopy was performed. STM topographs are 15 nm by 15 nm ($V$ = 10 mV, $I$ = 1 nA). A single CO molecule used for registration between spectra (*13*) accompanies each nanostructure. Spectral fingerprints ($dI/dV$ spectra) acquired throughout (**D**) Bilby, (**E**) Hawk, and (**F**) Broken Hawk. (**G**) The normalized averages of the Bilby and Hawk spectra match closely, consistent with isospectrality, while the average Broken Hawk spectrum differs significantly. Inset: Spectral correlation plot (dashed line denotes perfect match) quantifying Bilby-Hawk isospectrality and its departure in Broken Hawk.

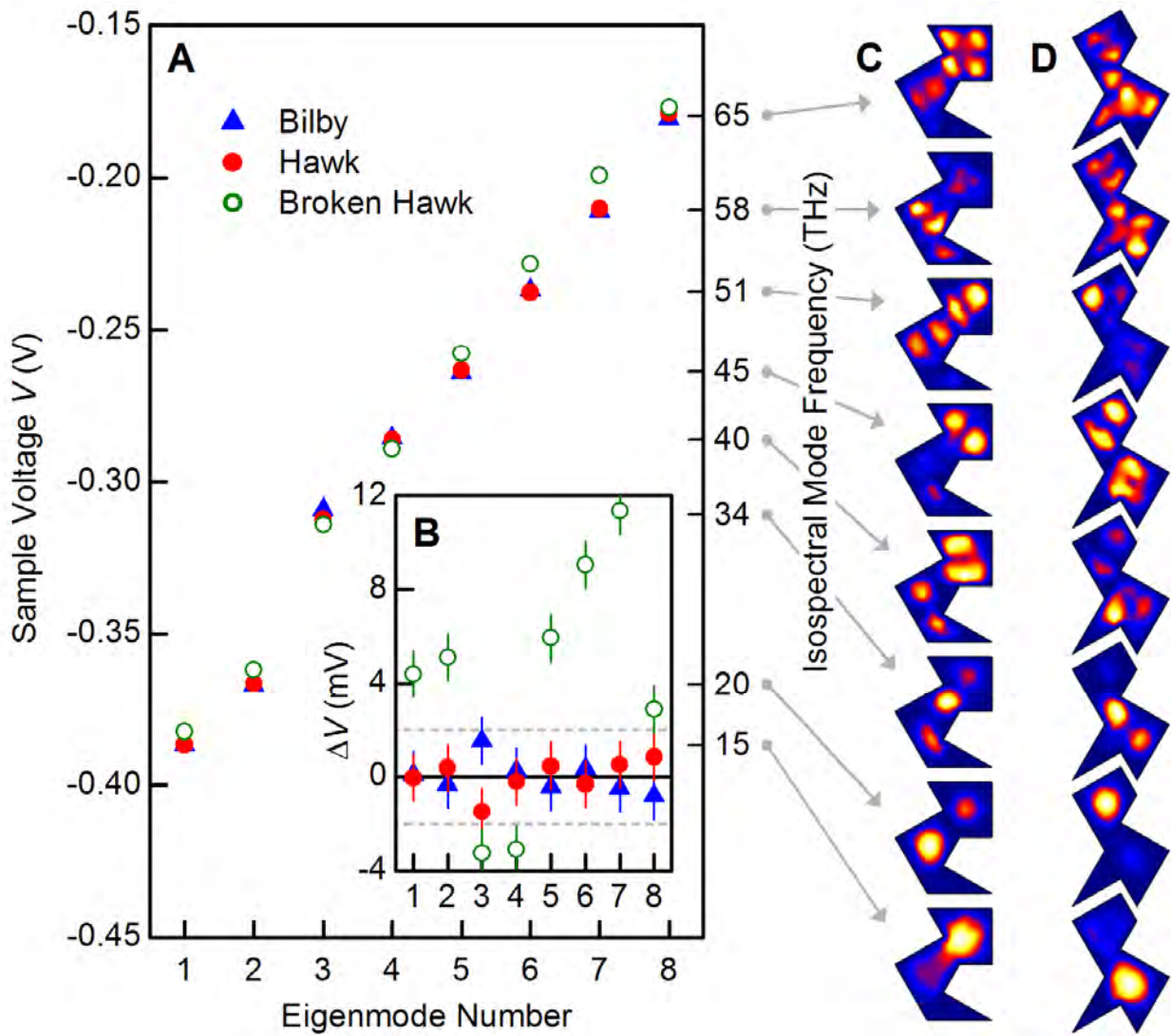

**Fig. 2 | Quantum isospectrality.** (**A**) The first 8 electron energy levels in (Δ) Bilby, (●) Hawk, and (○) Broken Hawk resonators, found by simultaneously fitting Lorentzians (see text) to all spectra in Fig. 1, D to F. (**B**) The difference between each of the energies in (A) and the mean of the Bilby-Hawk energy [error bars discussed in Methods (*13*)]. (**C** and **D**) Spatial maps of the amplitudes of the fitted Lorentzians reveal the eigenmodes of the isospectral structures Bilby and Hawk.

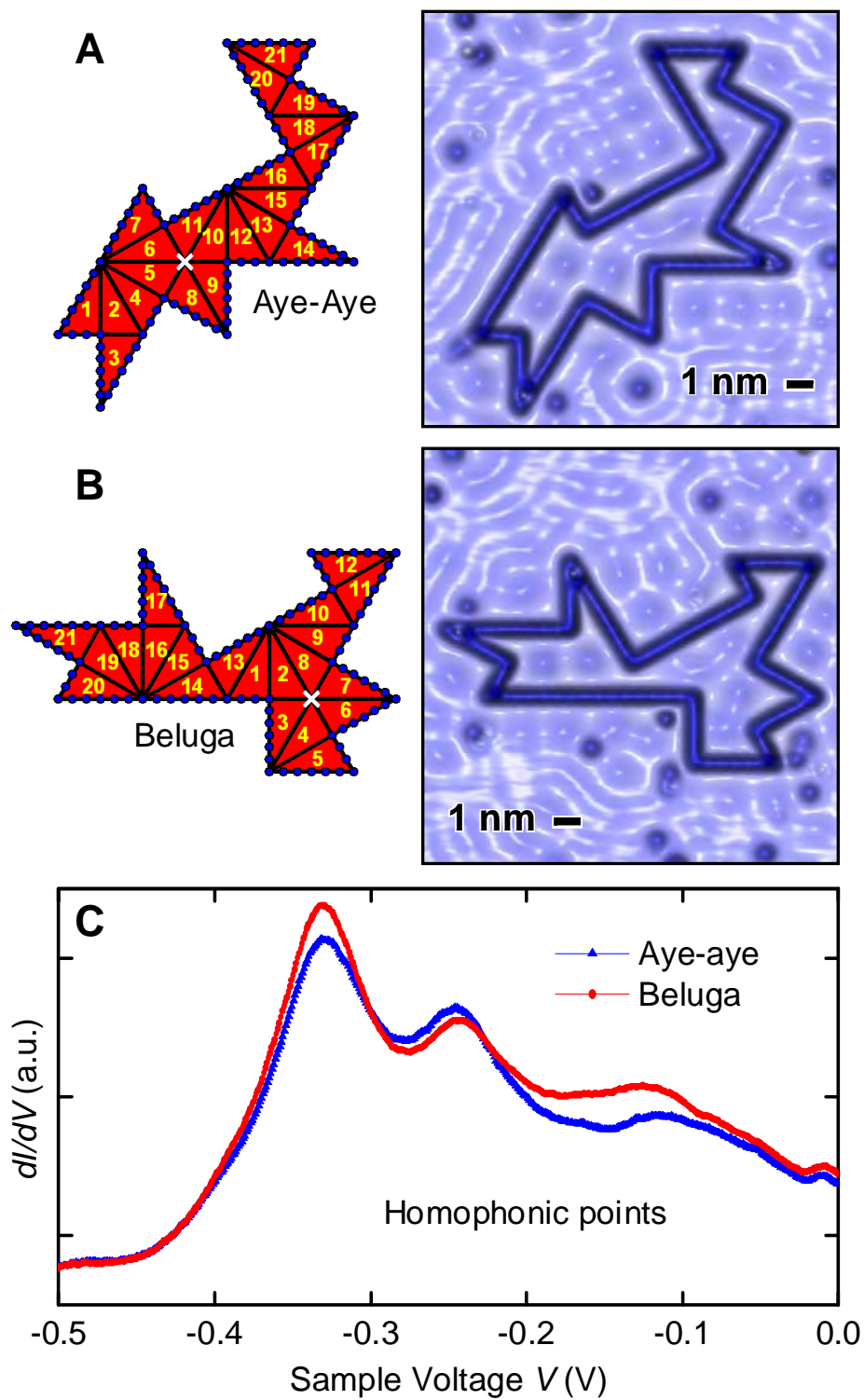

**Fig. 3 | Quantum homophonicity.** Designs and STM topographs for the (**A**) Aye-aye and (**B**) Beluga domains. Blue dots show molecular positions and white cross indicates the homophonic point of each structure. STM topographs ($V$ = 10 mV, $I$ = 1 nA) of $A$ (15 nm by 15 nm) and $B$ (16 nm by 16 nm) after assembly. (**C**) Spectra acquired at the two homophonic points are remarkably similar despite their entirely different environments.

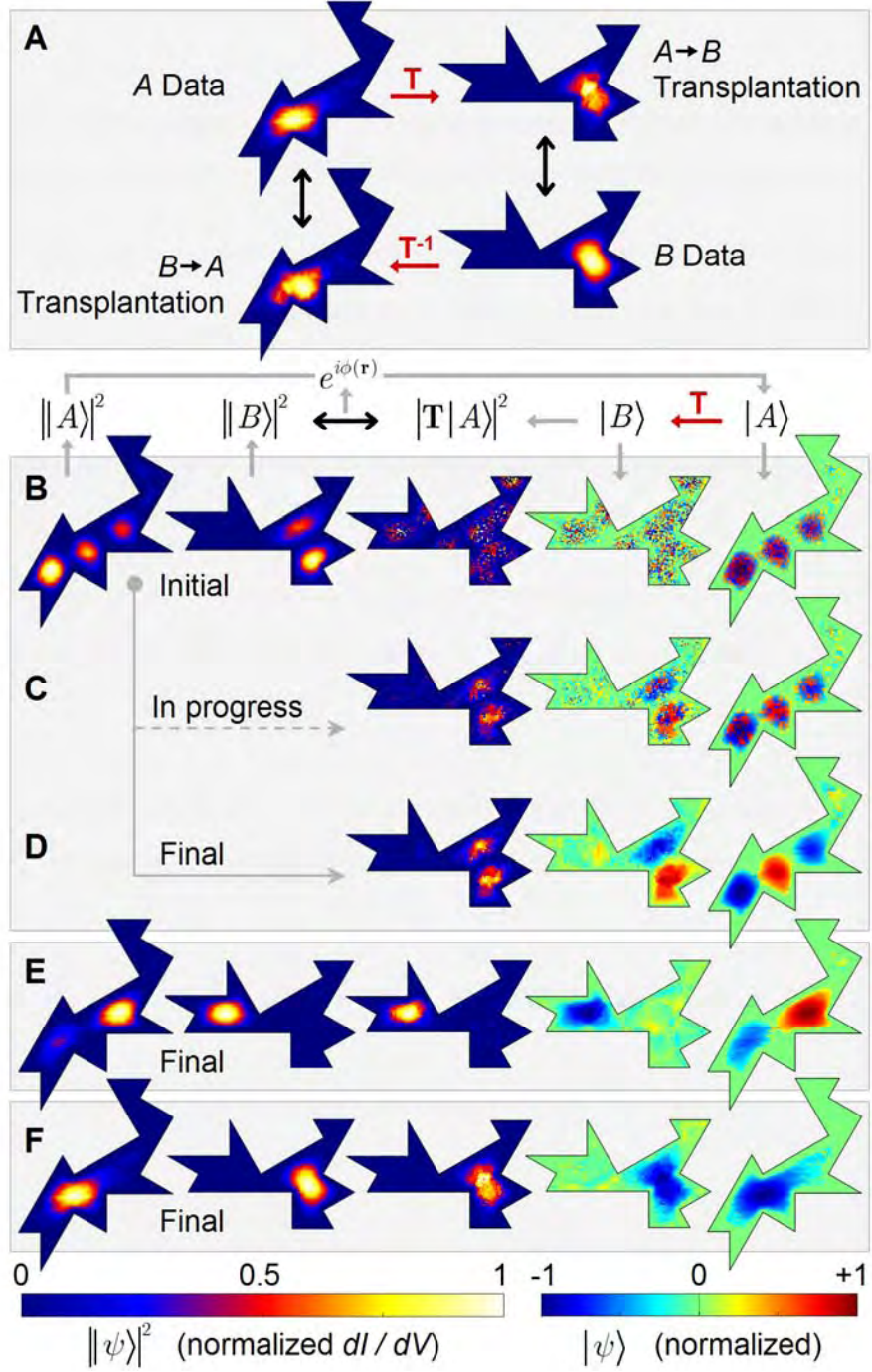

**Fig. 4.** (**A**) **Quantum transplantation.** Open-loop *dI/dV* map (*V* = -0.329 V) for the ground state of Aye-aye (upper left) is transformed by the transplantation operator **T** into a map that is an excellent match to the measured ground state eigenfunction for Beluga (lower right, same parameters). Similarly, **T**$^{-1}$ maps the *B* data onto the *A* shape, in excellent agreement with the measured *A* data. (**B** to **F**) **Quantum phase extraction.** (**B**) Initially, a random phase of 0 or $\pi$ is assigned to every data point, and the transplantation $|\mathbf{T}|A\rangle|^2$ is a poor match for $\||B\rangle\|^2$. (**C**) An intermediate state, with phase extraction still in progress. (**D**) The final extracted wave functions for the third eigenmode of the two shapes. (**E**) Results for the second mode. (**F**) The results for the first mode match those in (A) because all parts of the wave function are in phase. Colorbars normalized to maximum values.

– 13 –

# Quantum Phase Extraction in Isospectral Electronic Nanostructures

Christopher R. Moon,[1] Laila S. Mattos,[1] Brian K. Foster,[2] Gabriel Zeltzer,[3] Wonhee Ko,[3]
Hari C. Manoharan[1*]

[1]*Department of Physics, Stanford University, Stanford, CA 94305, USA*
[2]*Department of Electrical Engineering, Stanford University, Stanford, CA 94305, USA*
[3]*Department of Applied Physics, Stanford University, Stanford, CA 94305, USA*

*To whom correspondence should be addressed. E-mail: manoharan@stanford.edu

SUPPORTING ONLINE MATERIAL

**Supporting Text**

*Isospectrality and Transplantation*

While isospectrality and other facets of inverse spectral geometry (*S1, S2*) are a deep research topic in mathematics, the basic ideas of isospectrality are quite accessible and have been reviewed in several lay articles (*S3-6*). The proof of isospectrality in the shapes we discuss in the main paper is straightforward, and we will summarize it here. Fig. S1 shows the simplest pair of isospectral domains known, having only eight sides. These are the Bilby and Hawk studied in this work, but composed of seven isosceles right triangles rather than 30-60-90° triangles. The sides of each triangle have been colored differently; note that any two neighboring triangles are mirror images of one another (*S7*). Indeed, the following discussion can be connected to paper folding or origami, enabling "proof by folding" (*S8*).

Eigenmodes $\psi$ of the Bilby are solutions to the Helmholtz equation with Dirichlet boundary conditions; that is, $\nabla^2 \psi = -k^2 \psi$ inside the Bilby domain and $\psi = 0$ along its edges. To create eigenmodes of the Hawk, these wave functions are divided into triangular pieces that are linearly combined and transplanted into the Hawk triangles. Since the Laplacian is a linear operator, if these combinations of the Bilby pieces are smoothly continuous and vanish at the Hawk boundary, they must also be eigenfunctions of the Laplacian in the Hawk domain with the same eigenvalue $k$. Then, the two structures must be isospectral.

We label the seven Bilby triangles with the letters A–G. The Hawk triangles (Fig. S1 B) are numbered and labeled with a map that creates a Hawk eigenfunction from a Bilby eigenfunction. For instance, the map says that the top triangle of the Hawk (number 1) receives pieces A, E, and F from the Bilby. Because the E triangle is inverted with respect to this destination triangle (note the order of its sides), it needs to be "flipped over"; it is reflected about one of its sides and the wave function is inverted. In the map, it is marked with a minus sign. This map can also be represented as a 7 x 7 matrix that acts on a "column vector" of Bilby wave function triangles to create Hawk triangles. This is the representation used in the main text for the *A-B* isospectral shapes.

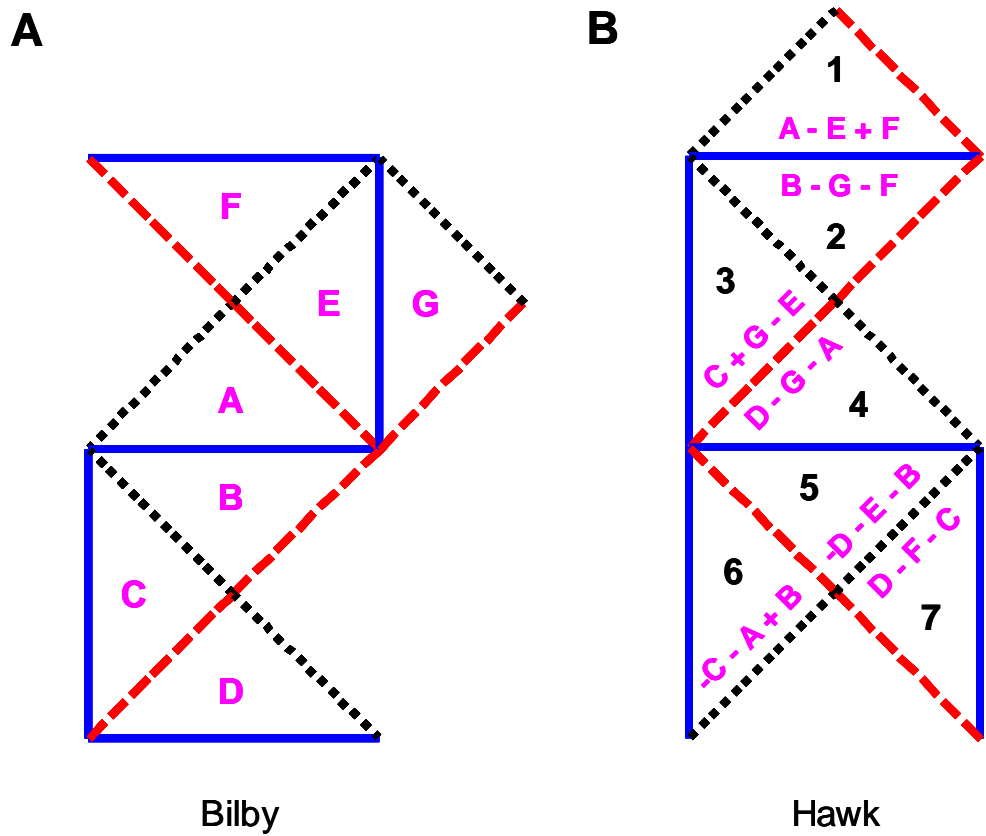

**Fig. S1 | The simplest pair of isospectral structures.** These domains are each tilings of seven 45-45-90° triangles, resulting in eight-sided polygons. (**A**) The Bilby, with triangles labeled arbitrarily. (**B**) The Hawk. Its triangles are numbered arbitrarily and labeled with a mapping that transforms any eigenfunction of the Bilby into its isospectral complement in the Hawk.

This map produces a valid Hawk eigenfunction because every relationship between pieces of a Bilby wave function is replicated in the Hawk. For example, a wave function in triangle A continues smoothly into B across the blue solid line. Accordingly, since A was pasted into triangle 1 in the Hawk, B has been transplanted into triangle 2. The Bilby A triangle continues into E across a red dashed line. However, the red dashed line of triangle 1 in the Hawk is along its border, so E was reflected backwards into triangle 1 and subtracted to assure that the resulting wave function goes smoothly to zero. Since every such relationship is satisfied, this map transforms all Hawk eigenfunctions into valid Bilby eigenfunctions.

What is the result if, as in our Broken Hawk (Fig. 1C), some of the triangular sections are flipped, breaking the reflection symmetry between neighboring triangles? This method for constructing transplantation maps then breaks down and isospectrality is no longer ensured. Experimentally, we observe significant changes to the spectra consistent with the breaking of isospectrality.

## Materials and Methods

*General*

The experiments described in this paper were repeated on many other isospectral structures of varying size and molecular wall densities, with confirmation of all results presented. We include in the article the data from the largest and most complex structures we have fabricated.

*Apparatus*

The experimental apparatus used for these studies was a home-built STM system. The STM sits in ultrahigh vacuum (UHV) ($< 10^{-11}$ Torr) and is located below four cascaded stages of passive vibration isolation on a low temperature stage thermally linked to a liquid helium reservoir. The entire STM system, consisting of a low-temperature chamber attached to room-temperature sample preparation and analysis vacuum chambers, is suspended above a separate subbasement foundation and surrounded by a soundproof acoustic chamber. The STM tip was polycrystalline Ir prepared by mechanical sharpening and in-situ field emission. The same tip configuration and sample conditions were preserved for all measurements performed on the Bilby/Hawk/Broken Hawk triplet, and the Aye-aye/Beluga pair.

*Sample Preparation and Molecular Manipulation*

The Cu(111) crystal was cleaned in UHV with several cycles of Ar sputtering and annealing at 600 °C before being lowered to the 4-K STM stage. CO molecules were dosed onto the sample by warming it to ~20 K and exposing the surface to 1 µTorr of CO gas for 30 seconds. The resulting coverage was approximately 10 molecules / (10 nm)$^2$ [the latter area being a typical image size] or ~0.012 monolayers. CO molecules were laterally manipulated by setting the tunnel junction resistance to ~250 kΩ. The walls of all resonators shown have been packed as tightly as possible with molecules ($2a$ in the [$10\bar{1}$] direction and $\sqrt{3}a$ in the [$1\bar{2}1$] direction, where $a = 0.255$ nm is the distance between adjacent Cu atoms at 4 K). The orbital overlap with neighbors causes the CO not to be individually resolvable along the perimeter.

*Data Acquisition*

Differential conductivity *dI/dV*, a probe of the local density of states, was acquired by injecting a small ac modulation ($V_{ac}$ = 4 mV rms at ~1 kHz) onto the dc sample bias *V* and measuring the response at the reference frequency using a lock-in amplifier. Each *dI/dV* spectrum was the average of 5 to 10 separate *dI/dV*-vs-*V* traces acquired at the same tip position. To maximize accuracy, the lateral and vertical (*z*) drift was first nulled to below 100 fm/min and then the following computer-controlled procedure was used to automate spectroscopy. Before each open-loop *dI/dV*-vs-*V* trace was acquired inside a nanostructure, the tip was registered to the center of an exterior reference molecule, compensating for any residual lateral drift. Then, the tip was moved to each point of interest in closed-loop (constant-*I*) mode at a relatively large bias (*V* = 0.5 V). Topographs at this voltage are homogeneous throughout the structures, ensuring a consistent tip height between spectra. Next, the feedback loop was opened and *dI/dV* was acquired as a function of *V*. The tip was then re-locked to the reference molecule and the cycle repeated until all traces at a given position were completed, and then the computer repeated this procedure at all programmed points of interest in a given nanostructure (point spacing was 4*a*). The data range examined encompasses the filled states of the 2D electron band, i.e. $-0.5\text{ V} \leq V \leq 0\text{ V}$, as the band begins at $V = -0.45\text{ V}$.

For open-loop *dI/dV* maps, first the alignment of the tip scanning plane was carefully adjusted to be parallel to the sample surface. Then the feedback loop was opened and the tip was rastered across the structure, thus keeping the tip at constant elevation. *dI/dV* was measured at each point with the same lock-in technique described above ($V_{ac}$ = 4 mV rms at ~6.5 kHz) while the sample dc bias *V* was held constant, resulting in a high-resolution image of local density of states vs. lateral position (*x*,*y*) at a particular sample voltage. The data is the sum of two tunneling pathways: from the tip to the sample's 2D surface states, which are confined into the eigenfunctions we aim to measure, and from the tip into the 3D bulk states. To obtain information solely due to the 2D electrons, we removed the latter contribution by simply subtracting a constant background and keeping the positive signal. In point spectroscopy, the 2D energy-dependent contribution to *dI/dV* is observed to ride above a constant background due to the 3D continuum, and this constant can be determined experimentally by looking at the step up in *dI/dV* at the 2D band edge at $V = -0.45\text{ V}$.

*Spectral Fitting Procedure*

In a given structure, we fit each of the acquired *dI/dV* spectra $S_j$ with *n* Lorentzians via $S_j(V) = c_j + \sum_{i=1}^{n}(a_{ij}w_i/2\pi)/\left[(V-V_i)^2 + (w_i/2)^2\right]$, where $V_i$ and $w_i$ are the energy and linewidth of the *i*th mode, $a_{ij}$ is the amplitude of the *i*th mode in the *j*th spectrum, $c_j$ is the *y*-offset of the *j*th spectrum, and *V* is the voltage of the sample with respect to the tip. Here, the number of Lorentzians *n* can be estimated with particle-in-a-box calculations, knowing the dispersion relation of the Cu(111) surface state band. In practice, varying *n* by several Lorentzians only changes the fit results for the high-energy ends of the spectra, as the data there is affected by modes outside the fitted energy domain; the fit results that change significantly when *n* is varied are discarded. An example of final fits for the entire Bilby data set is shown in Fig. S2. All structures were subjected to this ensemble fitting procedure with matching fidelity.

We find that this ensemble fitting procedure is highly constrained (each structure must produce only one set of mode energies $V_i$ and linewidths $w_i$ that simultaneously fit 46 *dI/dV* spectra, each containing 1024 voltage points), and thus leads to robust and low-error solutions ($\pm 0.1 \sim \pm 0.3$ mV for the Bilby, Hawk, and Broken Hawk data sets). Nevertheless, we also subject our fit results to several other checks such as (1) verifying the absolute values of the fit linewidths $w_i$ match known behavior, (2) verifying that the fit energies $V_i$ make sense with respect to numerical calculations of the energies via quantum scattering theory and particle-in-a-box models, and (3) verifying the nodal structure of each state. To accomplish this last task, the fitted amplitudes $a_{ij}$ are plotted as a function of interior position where the *j*th spectrum was acquired. This is shown for the first 8 modes ($i = 1...8$) for Bilby (Fig. 2C) and Hawk (Fig. 2D). We note that this method for imaging eigenmodes through fitted Lorentzian amplitude maps is complementary to direct maps of *dI/dV* at the mode energies; it provides maps devoid of contributions of neighboring states at the expense of data volume. Error bars (see Fig. 2B) for fit energies are determined from voltage measurement resolution, deviations in energy when number of Lorentzians *n* is varied (see above), and the confidence bounds of the fits. In Fig. 2A, error bars are smaller than the symbols.

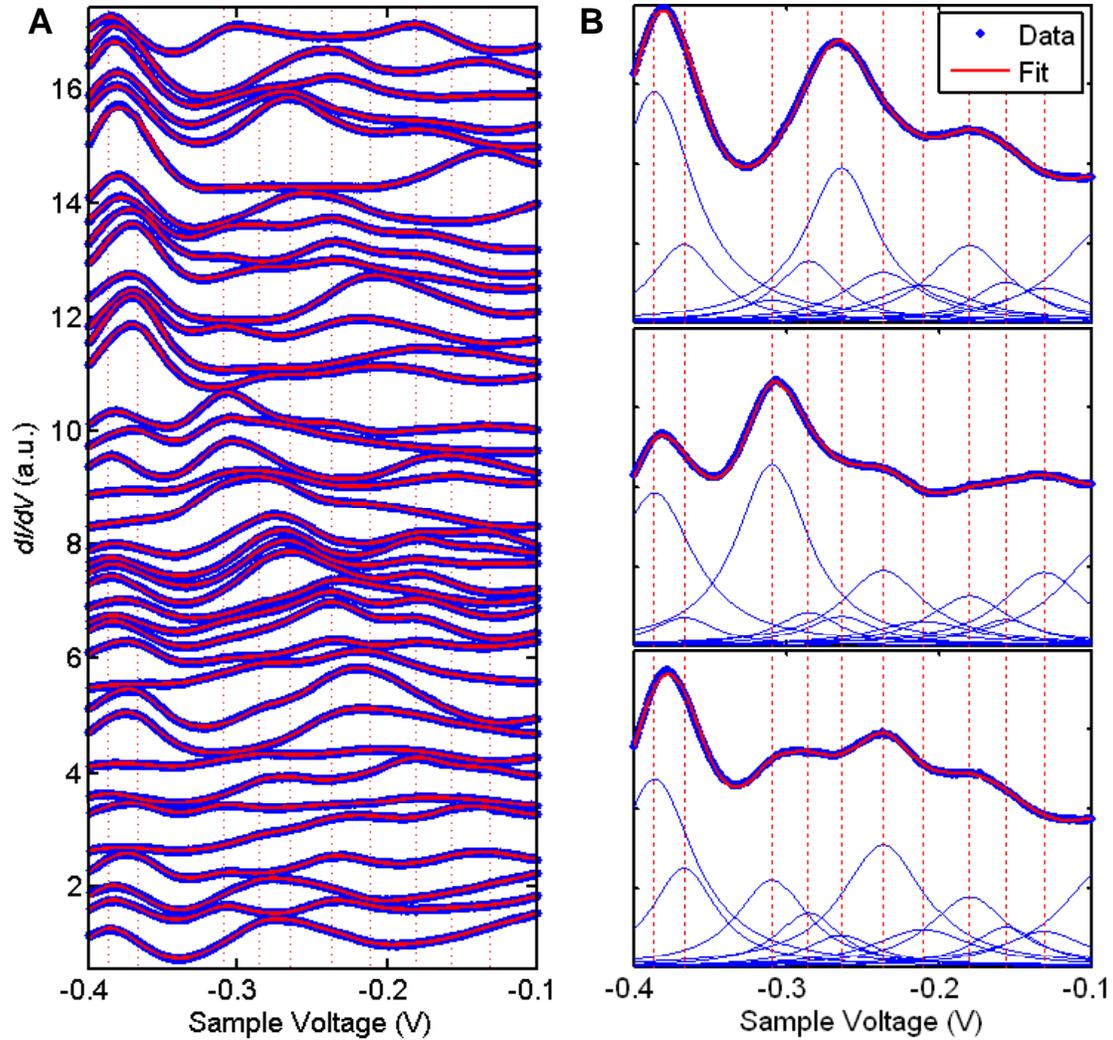

**Fig. S2 | Ensemble spectral fits.** (**A**) Measured traces (offset for clarity) of *dI/dV* vs. *V* (blue circles) acquired throughout the Bilby quantum resonator were simultaneously fit (solid red lines) with a set of 11 Lorentzians. Lorentzian centers, the fitted energies, are marked with vertical dashed lines. (**B**) Three sample traces from (A), additionally showing the individual Lorentzians (solid blue lines) which compose each fit. Only the Lorentzian amplitudes vary between spectra.

*Quantum Transplantation Machine (QTM)*

The transplantation matrix **T** that transforms wave functions of the *A* shape into wave functions of the *B* shape (see main text) is given in Eq. S1 below.

$$\mathbf{T} = \begin{pmatrix}
0 & 0 & 0 & 0 & 1 & 0 & 1 & -1 & 0 & 0 & 0 & 0 & 0 & 0 & 1 & 0 & 0 & -1 & 0 & 0 & 0 \\
0 & 0 & 0 & 1 & 0 & 1 & 0 & 1 & 0 & 0 & 0 & 0 & 1 & 0 & 0 & 0 & -1 & 0 & 0 & 0 & 0 \\
0 & 0 & 0 & -1 & 1 & 0 & 0 & 0 & 0 & 1 & 0 & 0 & 1 & 0 & -1 & 0 & 0 & 0 & 0 & 0 & 0 \\
0 & -1 & 0 & 0 & 0 & 1 & 0 & 0 & 0 & 1 & 0 & 0 & 0 & -1 & -1 & 0 & 0 & 0 & 0 & 0 & 0 \\
-1 & 0 & 0 & 0 & 0 & 0 & 1 & 0 & 0 & 0 & 0 & 1 & -1 & 1 & 0 & 0 & 0 & 0 & 0 & 0 & 0 \\
0 & 0 & -1 & 0 & 0 & 0 & 0 & 0 & 1 & 0 & 1 & 0 & -1 & 0 & 1 & 0 & 0 & 0 & 0 & 0 & 0 \\
0 & 0 & 1 & 0 & 0 & 0 & 0 & 1 & 0 & 1 & 0 & -1 & 0 & 0 & 0 & 1 & 0 & 0 & 0 & 0 & 0 \\
0 & 1 & 0 & 0 & 1 & 0 & 0 & 0 & 1 & 0 & 0 & 1 & 0 & 0 & 0 & 0 & 1 & 0 & 0 & 0 & 0 \\
1 & 0 & 0 & 1 & 0 & 0 & 0 & 0 & -1 & 1 & 0 & 0 & 0 & 0 & 0 & 0 & 0 & 1 & 0 & 0 & 0 \\
-1 & 1 & 0 & 0 & 0 & 0 & 0 & -1 & 0 & 0 & 1 & 0 & 0 & 0 & 0 & 0 & 0 & 0 & 1 & 0 & 0 \\
1 & 0 & 1 & 0 & -1 & 1 & 0 & 0 & 0 & 0 & 0 & 0 & 0 & 0 & 0 & 0 & 0 & 0 & 0 & 1 & 0 \\
0 & 1 & -1 & -1 & 0 & 0 & 1 & 0 & 0 & 0 & 0 & 0 & 0 & 0 & 0 & 0 & 0 & 0 & 0 & 0 & 1 \\
0 & 0 & 0 & 0 & 0 & 1 & -1 & 0 & -1 & 0 & 0 & 0 & 0 & 0 & 1 & 0 & 0 & -1 & 0 & 0 \\
0 & 0 & 0 & 0 & 0 & 0 & 1 & 0 & 0 & -1 & 1 & 0 & 0 & 0 & 0 & 1 & 0 & 0 & -1 & 0 \\
0 & 0 & 0 & 0 & 0 & 1 & 0 & 0 & 0 & 0 & -1 & -1 & 0 & 0 & 0 & 0 & 1 & 0 & 0 & -1 \\
0 & 0 & 0 & 0 & 1 & 0 & 0 & 0 & 0 & -1 & 0 & 0 & -1 & 0 & 0 & 0 & 0 & 1 & 0 & 1 \\
0 & 0 & 0 & 0 & 0 & 0 & 0 & 1 & -1 & 0 & 0 & 0 & 0 & -1 & 0 & 0 & 0 & 0 & 0 & 1 & -1 \\
0 & 0 & 0 & 1 & 0 & 0 & 0 & 0 & 0 & 0 & -1 & 0 & 0 & -1 & 0 & 0 & 0 & -1 & 1 & 0 \\
0 & 1 & 0 & 0 & 0 & 0 & 0 & 0 & 0 & 0 & 0 & -1 & 0 & 0 & -1 & 0 & -1 & 0 & -1 & 0 \\
1 & 0 & 0 & 0 & 0 & 0 & 0 & 0 & 0 & 0 & 0 & 0 & 0 & -1 & 1 & -1 & 0 & 0 & 0 & -1 \\
0 & 0 & 1 & 0 & 0 & 0 & 0 & 0 & 0 & 0 & 0 & 0 & 0 & -1 & 0 & 0 & -1 & 1 & -1 & 0 & 0
\end{pmatrix} \quad (S1)$$

The phase extraction method using this operator is shown in Fig. S3. The following is a description of the algorithm in the "phase tuner" box. We applied a greedy algorithm (*S9*): the signs (corresponding to phases 0 and $\pi$) were flipped individually in random order, and if a flip decreased the total error $\delta$, it was accepted. This procedure was continued until a local minimum was reached. Then, we applied a relaxation step in which points whose phase was different from the majority of its nearest neighbors were flipped. This caused the total error to increase, but allowed subsequent greedy phases to reach a lower local minimum. This process continued until the system converged on what was assumed to be the global minimum, and the algorithm terminated when $\delta < \epsilon$, input as a convergence criterion. We also found that the specifics of the algorithm were not critical. For example, an alternate algorithm based on simulated annealing also converged to the same phase-extracted wave functions.

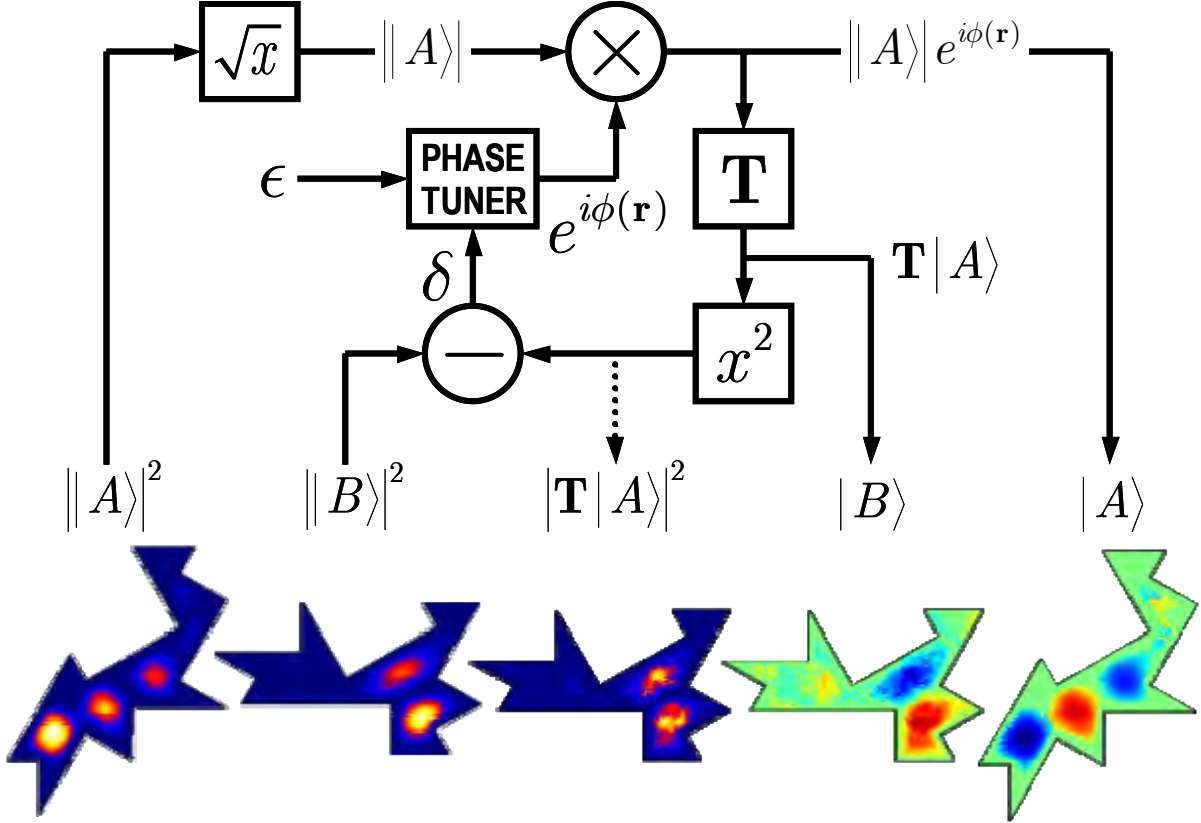

**Fig. S3 | Schematic of quantum transplantation machine.** Given density of states data for a mode in shape *A* (first column), the algorithm determines the quantum-mechanical phase of the corresponding wave function $|A\rangle$ such that its transplantation (third column) matches the data for its isospectral partner in shape *B* (second column). The outputs are the full wave functions $|A\rangle$ (fifth column) and $|B\rangle$ (fourth column). Colorbars as in Fig. 4.

    For optimized extraction, we accounted for the fact a *dI/dV* map at one eigenenergy will contain some contributions from the neighboring modes. We corrected for this spurious signal to first order by subtracting a small proportion of maps $i\pm1$ from map $i$, such that map $i$ remained strictly positive but had its contributions from neighboring modes greatly reduced. A further check of this procedure was to match to experimental Lorentzian amplitude images (as in Fig. 2, C and D, with analysis applied to homophonic structures) which provide mode maps without contributions from neighboring modes. After this correction and background subtraction, our *dI/dV* maps are considered images of $|\psi|^2$, the magnitude-squared wave functions of the isospectral structures. We also verified that the QTM provided correct phase extraction, accompanied by more noise, on uncorrected data. Even though in all cases the final output could be made nearly perfect in appearance by performing a final local average or smoothing (for example this would remove lone phase-inverted pixels), this step was omitted to keep the QTM as simple as possible and true to the task of returning a minimum error solution for the wave functions.

The success of transplantation itself is a strong justification of these methods. However, we have also checked the validity of the extracted wave functions by confirming their mutual orthogonality after each wave function had been properly normalized. We also confirmed successful operation of the QTM in Bilby and Hawk by operating on the lower-resolution state maps shown in Fig. 2, C and D.

*Sound Conversion*

For illustrative purposes and because of the strong analogy to the classical drum problem, we converted the *dI/dV* spectra of our "quantum drums" into audible sound. First, the electron energy *E* was converted to frequency $\nu$ via $h\nu = E - E_0$, where $E_0 = -0.45$ eV is the band edge for surface state electrons on Cu(111) and *h* is Planck's constant. The resulting energy scale, shown in Figs. 1 and 2, represents the natural electron resonance frequencies in these nanostructures, with the Fermi energy corresponding to ~100 THz. To map the real electron dynamics onto the human audible range (20~20,000 Hz), we scaled $\nu$ by a factor of $10^{-11}$. Then to compensate for the logarithmic response of the human ear, we exponentiated the data such that the measured spectrum mapped roughly to a decibel scale [$(dI/dV)^{10}$]. Finally, to avoid chirped signals, we added a random phase to each frequency component as is common in audio processing for spectral conversion. We then took the inverse Fourier transform to create the sounds. The converted spectral bandwidth (sampling rate in time domain) was 44.1 kHz, with 1 Hz resolution and duration of 1 sec. The percussive nature of the resulting sounds is a natural product of the spectral content and the inverse Fourier transform which creates a falling and rising amplitude envelope. We take the unique part of the final time series record (the first half, duration 0.5 sec) and repeat it several times to "listen" to the assembled quantum drums.

**Movie Captions**

*Movie S1 – Isospectrality (QuickTime, 1.7 MB, video and audio track)*

Sounds generated (see Methods) from the average spectra (Fig. 1G) acquired inside the Bilby, Hawk, and Broken Hawk shapes. Each sound is repeated 8 times as the topograph of the corresponding quantum drum is highlighted. Bilby and Hawk—which are isospectral—"sound the same," while Broken Hawk can be audibly distinguished.

*Movie S2 – Homophonicity (QuickTime, 3.7 MB, video and audio track)*

Audio conversion (see Methods) of 3 pairs of point spectra acquired within the isospectral Aye-aye and Beluga shapes, including the homophonic pair. The similarity between the homophonic points is audibly contrasted with the easily discernable differences between other points. Each sound is repeated 4 times as the topograph of the corresponding nanostructure is highlighted and the position of the acquired spectrum is identified with a red cross.

*Movie S3 – Transplantation (QuickTime, 15.7 MB, video and audio track)*

Full phase extraction process for modes 1 to 3 of the Aye-aye (*A*) and Beluga (*B*) quantum resonators. The algorithm is represented by a black box, the quantum transplantation machine (QTM), which has the goal of obtaining the wave function $\psi(\mathbf{r}) = |A\rangle$ from the measured probability density $P(\mathbf{r}) = |\psi|^2 = \||A\rangle\|^2$. The raw inputs to the QTM (at left) are the measured probability densities $|\psi|^2$ of an isospectral state pair in *A* and *B*. The outputs (at right) are the full wave functions $\psi$ of the two shapes. The algorithm (center panels) determines the phase $\phi(\mathbf{r})$ of the *A* mode by optimizing the match between its transplantation **T** and the measured *B* density (the "feedback"). In the first frame, $\phi(\mathbf{r})$ is random. Subsequent frames are recorded after each of two operations: (1) the phase of every pixel is varied in random order, with each change accepted if the total error (red bar) decreased; (2) points whose phase differed from the majority of their neighbors were inverted. The second step tends to increase total error momentarily while allowing subsequent iterations to find a lower minimum. Chapter headings, matching audio narration, can be selected to jump to specific QTM states.

**Supplementary Figures**

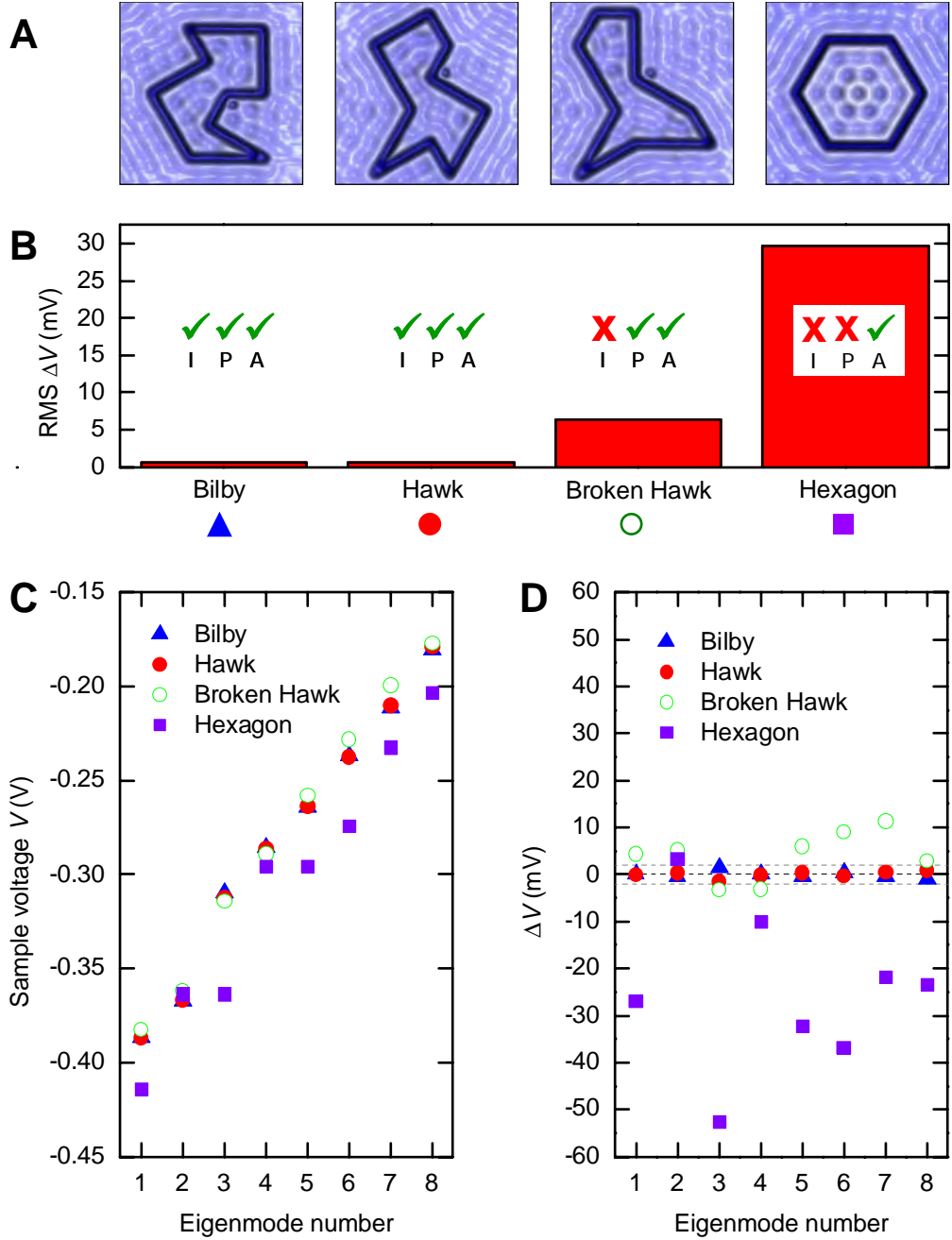

**Fig. S4 | Breaking isospectrality.** (**A**) Topographs of 4 electron resonators (15 x 15 nm; 10 mV, 1 nA; color scale as in Fig. 1). (**B**) The RMS differences of the first 8 eigenenergies of each structure from the average Bilby-Hawk energies show the net effect of altering isospectral geometry ("I") and perimeter ("P") in corrals of identical area ("A"). (**C**) The electron energy levels in these quantum corrals were measured via spectral mapping. (**D**) As discussed in the text, the Bilby and Hawk shapes are isospectral to within ± 2 mV. In Broken Hawk, an underlying tiling symmetry is violated but area and perimeter are preserved. In the hexagon, the perimeter must change for constant area, which greatly impacts the spectrum. (Because a hexagon cannot be constructed with the exact same area, the energies shown for that shape have been scaled slightly to account for this).

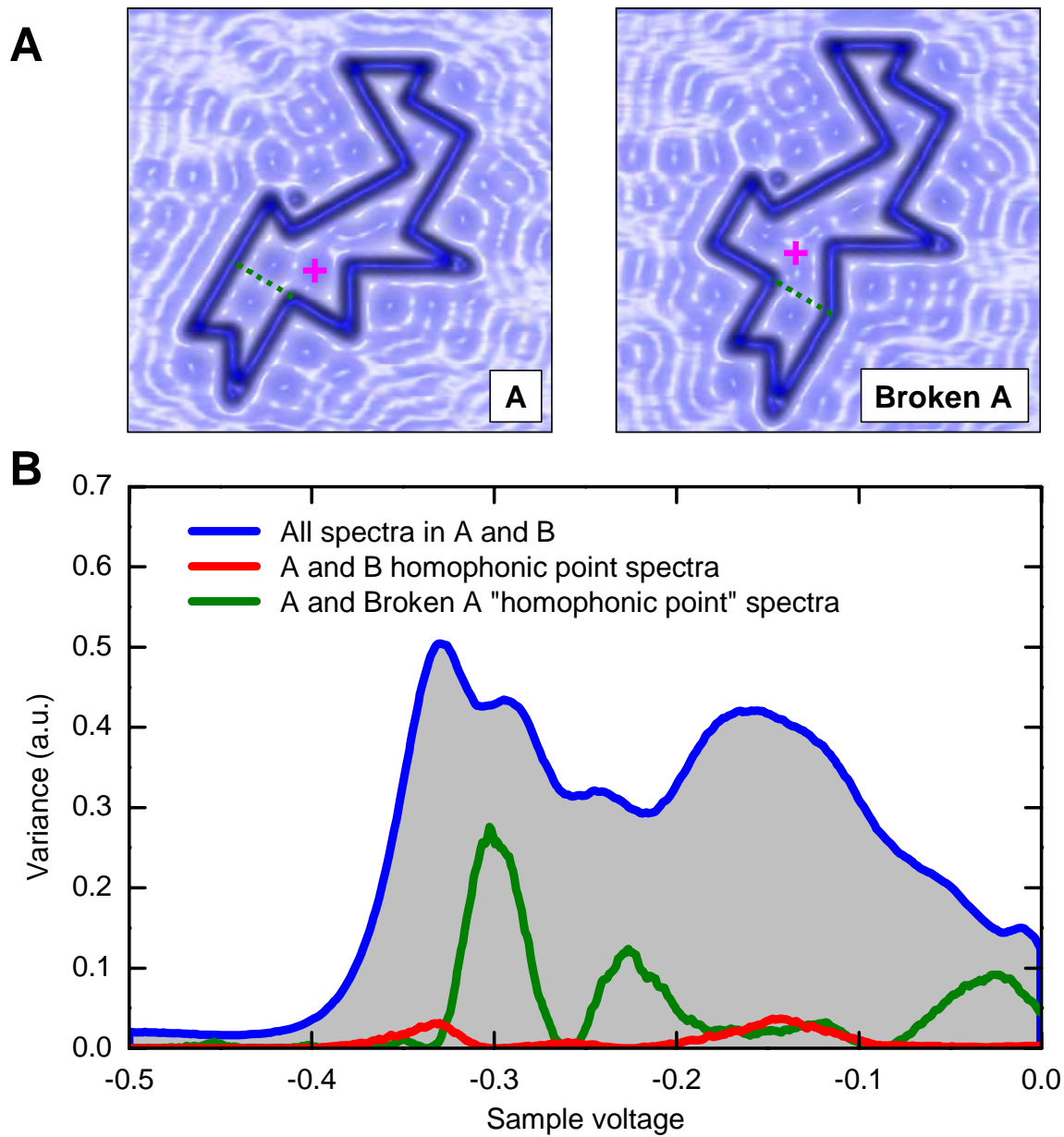

**Fig. S5 | Breaking isospectrality and homophonicity.** (**A**) Topographs of an *A*-shaped quantum corral and its broken version (17 nm x 17 nm; 10 mV, 1 nA). The Broken *A* was created by shifting the section demarcated by the green dotted line. Crosses mark the homophonic point in *A* and the same point in Broken *A*. (**B**) The variance of all spectra taken throughout the *A*- and *B*-shaped electron resonators (blue line) is far greater than that of the two homophonic traces (red line, showing variance between the two traces of Fig. 3C). When the *A* shape is altered slightly, breaking isospectral geometry, this causes far more variation (green line) in the homophonic point spectra than when the altered shape was the radically different *B* domain (in which isospectral geometry was preserved, Fig. 3, A and B).

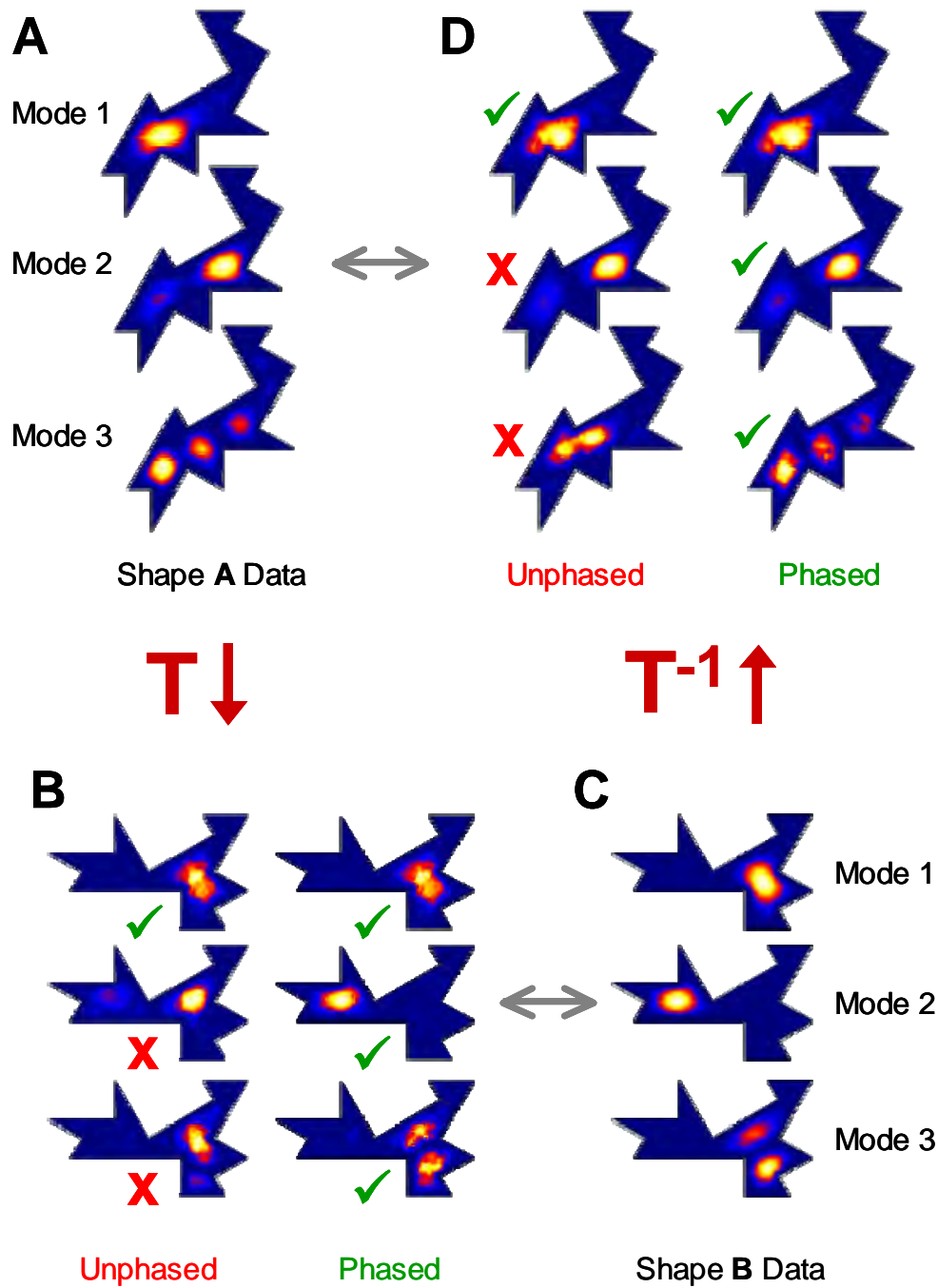

**Fig. S6 | The necessity of quantum phase in transplantation.** (**A**) The first three eigenmode probability densities $|\psi|^2$ of the *A*-shaped quantum corral, as determined from *dI/dV* maps. If the wave functions $\psi$ are transplanted onto the *B* shape without accounting for quantum mechanical phase (**B**, first column), the results for the excited states are markedly different from the measured *B* eigenmode data (**C**). When the phase is extracted using the methods presented in this work, the transplantations (**B**, second column) match the data well. The same is true for the inverse: the *B* data also requires phase extraction to be properly transplanted onto the *A* shape (**D**). Only the ground states transplant correctly without phase adjustment because their wave functions are entirely in phase.

**Supporting References and Notes**